\documentclass[conference]{IEEEtran}
\IEEEoverridecommandlockouts

\usepackage{cite}
\usepackage{amsmath,amssymb,amsfonts}
\usepackage{algorithmic}
\usepackage{graphicx}
\usepackage{textcomp}
\usepackage{xcolor}
\usepackage{booktabs}
\usepackage{array}
\usepackage{multirow}
\usepackage{url}
\usepackage{placeins}
\usepackage{capt-of}

\def\BibTeX{{\rm B\kern-.05em{\sc i\kern-.025em b}\kern-.08em
T\kern-.1667em\lower.7ex\hbox{E}\kern-.125emX}}

\newcommand{\paperfig}[4]{%
\begin{figure}[!htbp]
\centering
\includegraphics[width=#1\columnwidth]{#2}
\vspace{-1mm}
{\footnotesize
\caption{#3}
\label{#4}}
\vspace{-1mm}
\end{figure}
}

\begin{document}

\title{Evaluating Conformal Reliability of Pathway-Level Transcriptomic Signatures Under Cross-Cohort Shift in Sepsis Mortality Prediction}
\author{
\IEEEauthorblockN{Pratyush Kumar Shukla}
\IEEEauthorblockA{
\textit{Independent Researcher}\\
Frisco, TX, USA\\
pratyushs2009@gmail.com
}
\and
\IEEEauthorblockN{Manveer Singh Tib}
\IEEEauthorblockA{
\textit{Independent Researcher}\\
Frisco, TX, USA\\
Mnvertib@gmail.com
}
\and
\IEEEauthorblockN{Siddhant Garg}
\IEEEauthorblockA{
\textit{Independent Researcher}\\
Frisco, TX, USA\\
sidg0379@gmail.com
}
}

\IEEEaftertitletext{\vspace{-2.5\baselineskip}}

\maketitle

\begin{abstract}
\normalfont
Blood transcriptomic profiling enables prognostic modeling by capturing the host immune response at the molecular level. Yet, the within-cohort evaluation strategies employed by many transcriptomic models inadequately reflect deployment across independent hospitals. Outside deployment scenarios introduce a cohort shift that can substantially degrade predictive performance and reliability of uncertainty estimates. We present a framework for evaluating transcriptomic sepsis mortality prediction under realistic cross-cohort deployment, systematically comparing gene-level, pathway-level and hybrid molecular representations. Four publicly available whole-blood transcriptomic cohorts consisting of 936 patients and 248 mortality events were harmonized into a shared 7,660-gene feature space and evaluated under leave-one-cohort-out validation using logistic regression, random forests, XGBoost and LightGBM. Beyond AUROC and AUPRC, model behavior was evaluated via conformal prediction, calibration analysis, selective prediction and the proposed Pathway Stability Index. Gene-level and hybrid representations were found to generally achieve the strongest discriminative performance, whereas pathway-level representations exhibited greater robustness across model families, more reliable uncertainty behavior under cross-cohort shift and stable molecular signatures enriched for immune and host-defense processes identified through Gene Ontology and KEGG enrichment analyses. These findings demonstrate that molecular representation influences not only predictive discrimination but also calibration, uncertainty reliability, biological coherence and transferability under external validation.
\end{abstract}


\section{Introduction}

Sepsis, a leading cause of death among critically ill patients worldwide, is a life-threatening organ dysfunction arising from a dysregulated host response to infection [1]. Given the progressive nature of the illness, clinical importance of early mortality identification is crucial to guide treatment escalation, inform ICU monitoring decisions and support risk stratification while faced with a heterogeneous disease. In light of this, comprehensive analysis of RNA molecules within a blood sample, blood transcriptomic profiling (BTP), has emerged as a viable method for mortality prediction by leveraging molecular-level responses of host immune systems. An increasing array of literature has thus focused on the ability to ascertain biomarkers and predict sepsis outcomes through the use of machine learning models utilizing transcriptomic measurements.

In this space, various transcriptomic prediction models have achieved encouraging results when evaluated with random train-test splits or cross-validation within a single cohort. Though, these training regimes do not adequately reflect the conditions imposed by real-world deployment scenarios. For the purpose of clinical adoption, a model trained on data from one institution or study must exhibit generalizability to patient data collected across a multitude of other clinical workflows. Within sepsis transcriptomics, differentiations regarding patient metrics and clinical practice result in a significant cohort shift. As a consequence of this transition, models that exhibited strength under pooled or single cohort evaluation may fail when transferred to an external cohort. Traditional discrimination assessment metrics, notably the area under the receiver operating characteristic curve (AUROC) and area under the precision-recall curve (AUPRC), provide relatively little information about model confidence and trustworthiness under shifting patient cohorts: underscoring a primary concern for models required to perform optimally within clinical decision support tooling. Cautious models that acknowledge uncertainty and/or contain embedded abstention processes may be therefore preferred against overconfident and incorrect predictions.

Choosing appropriate molecular representations poses another challenge. Resolution and dimensional disparities between gene and pathway-level model training are apparent: Models trained on the gene-level boast maximized transcriptomic resolution retention and the ability to capture detailed predictive signals. Vulnerability to overfitting or cohort-specific noise accompanied with high dimensionality are however inherent negatives. Pathway-level representations offer biologically meaningful contextual aggregation of genes, including inflammatory signaling, immune activation or cell-cycle processes. This aggregative approach at the pathway level is able to reduce dimensionality but may sacrifice on bare predictive resolution. Third, hybrid representations that combine genes and pathways may balance precise signals in tandem to existing biological structure. Despite pathway scoring methods such as GSVA and ssGSEA featuring widespread usage within transcriptomic analysis, comparatively few studies have systematically contrasted gene-level, pathway-level and hybrid representations across cohorts while simultaneously assessing formalized key metrics.

We propose the idea of pathway-level uncertainty and stability for external reliability to address these disparities. Our framework is built specifically to evaluate transcriptomic sepsis mortality prediction under realistic cohort shifts. Four independent sepsis cohorts were assembled for this task, spanning 936 patients and 248 deaths total. This data was then aligned into a common feature space containing 7,660 genes and three complementary representations of gene-level expression, Hallmark pathway activity scores and a hybrid gene-and-pathway representation were constructed. Each individual representation was afterwards assessed under leave-one-cohort-out (LOCO) validation using logistic regression, random forest, XGBoost and LightGBM. We therefore evaluated not only raw discrimination, but also conformal reliability, calibration, abstention behavior and feature stability. A Pathway Stability Index (PSI) based upon resampling and directional reproducibility analyses was also introduced for quantifying stable transferable signals.

Primary contributions made in this work:
\begin{itemize}
\setlength{\itemsep}{1pt}
\setlength{\parskip}{0pt}
\setlength{\parsep}{0pt}
\setlength{\leftskip}{0pt}
    \item Production of a four-cohort benchmark for sepsis transcriptomic prognosis across cohorts: 936-patient mortality prediction benchmark assembled across GAinS, GSE65682, GSE95233 and GSE54514. This structure enables LOCO evaluation under realistic hospital shift.
    \item Systematic comparison of molecular representations under external validation: To determine how representation choice affects generalization, gene-level, pathway-level and hybrid transcriptomic representations were compared across logistic regression, random forest, XGBoost and LightGBM.
    \item Evaluation of transcriptomic prognostic models focused on reliability: Conformal coverage, abstention behavior, calibration and model-choice robustness were evaluated in addition to AUROC and AUPRC. The question at hand remained whether uncertainty estimates remain reliable on excluded cohorts and a keen focus was placed on this area.
    \item Development of a stable transferable biology identification framework: PSI was employed to classify stable transcriptomic signals. We illustrated the fact that high-stability pathways were more reproducible across cohorts and enriched for immune and host-defense processes relevant to sepsis pathophysiology.
\end{itemize}

The analyses indicate that deployment-relevant model behavior under cross-cohort shift is influenced by representation choice. Compared with pathway-level models, gene-level and hybrid models may achieve stronger raw discrimination under some settings; however, pathway-level representations can provide a lower-dimensional and reliability-favoring trade-off with improved stability, uncertainty behavior and biological coherence. The central recommendation is therefore not that one representation is universally best, but that transcriptomic prognostic models should be evaluated jointly for discrimination, calibration, uncertainty reliability and biological reproducibility.

\section{Related Work}

\subsection{Sepsis Transcriptomic Prognostic Modeling}

Previous work has shown that blood gene-expression profiles contain prognostic information regarding sepsis outcomes. Sweeney et al. developed transcriptomic models for sepsis mortality using a community modeling strategy across different publicly available cohorts. This demonstrated that it is possible for transcriptomic signals to support mortality risk prediction beyond a single study population [2]. Subsequent work has continued to explore immune-related transcriptomic signatures for sepsis prognosis. This has included immune paralysis signatures and other host response gene sets associated with 30-day mortality [3]. Machine learning approaches have also been applied to identify gene-expression biomarkers that are associated with complicated sepsis courses. One example is pediatric sepsis cohorts where models derived from gene-expression were evaluated using cross-validation strategies [4].

\subsection{Pathway-Level Transcriptomic Representations}

Despite the fact that gene-level transcriptomic models preserve high-resolution molecular information, their dimensionality can make them vulnerable to noise, overfitting and other platform effects. Pathway-level representations give us an alternative as they aggregate genes into biologically interpretable programs. Gene Set Variation Analysis (GSVA) was introduced as an unsupervised method for estimating variation in pathway activity across samples from microarray or RNA-seq expression data [5]. The Hallmark gene sets from the Molecular Signatures Database (MSigDB) provide a curated collection of 50 refined biological signatures designed to reduce redundancy and represent coherent biological states or processes [7]. 

\subsection{Uncertainty and Conformal Prediction Under Distribution Shift}

Within clinical machine learning, predictive uncertainty is often assessed to be just as important as point prediction. Conformal prediction allows for a model-agnostic framework for producing prediction sets or intervals with user-specified coverage guarantees under exchangeability assumptions [8], [9]. Modern conformal methods have become attractive due to their ability to wrap around black-box predictors and provide uncertainty sets with finite-sample and distribution-free guarantees in situations where calibration and test data are exchangeable [9]. Though, this assumption is challenged in biomedical deployment settings where the test cohort may differ substantially from the calibration cohort. Work on conformal prediction under covariate shift explicitly shows that standard conformal guarantees may require adaption when training and test covariate distributions differ [10]. Open implementations like MAPIE have made conformal methods easier to apply in the context of regression and classification settings. Yet, it is also true that the reliability of conformal coverage still depends on whether the calibration data is representative of the deployment distribution [11]. A central deployment question raised for sepsis transcriptomic prognosis is then raised: when a model is trained and calibrated on some cohorts but evaluated on a new hospital or study, does its uncertainty remain trustworthy? The work we completed addresses this gap by evaluating gene-level, pathway-level and hybrid transcriptomic representations by conformal coverage; abstention behavior; calibration; and stability under LOCO shift instead of solely AUROC and AUPRC.

\section{Data and Cohort Construction}

\subsection{Cohort Selection}

Data was obtained and assembled from four publicly available transcriptomic cohorts that contained mortality outcomes for patients suffering from sepsis. To accurately evaluate model performance under realistic deployment conditions, cohorts were sourced from independent studies conducted at different institutions. We ensured that cohort data was also generated via distinct patient populations and expression platforms. The final benchmark consisted of the GAinS cohort [12] together with the GEO datasets GSE65682 [13], GSE95233 [14] and GSE54514 [15] giving a combined total of 936 patients including 248 deaths (Table~\ref{tab:cohorts}).

Each cohort was chosen to be kept as an independent unit throughout our analysis as opposed to utilizing sample pooling. The rationale was that independence allowed for LOCO evaluation where one hospital or study was excluded during training and reserved exclusively for external testing. Resulting from this, each reported performance metric reflects generalization to a previously unseen cohort versus interpolation within a mixed dataset. Table~\ref{tab:cohorts} summarizes the cohorts included in the benchmark together with their mortality distributions.

\begin{table}[t]
\centering
\caption{Summary of sepsis transcriptomic cohorts used in this study.}
\label{tab:cohorts}
\small
\setlength{\tabcolsep}{4pt}
\renewcommand{\arraystretch}{1.10}
\begin{tabular}{@{}lcccc@{}}
\toprule
\textbf{Cohort} & \textbf{Source} & \textbf{Patients} & \textbf{Survivors} & \textbf{Deaths} \tabularnewline
\midrule
GAinS & E-MTAB & 371 & 263 & 108 \tabularnewline
GSE65682 & GEO & 479 & 365 & 114 \tabularnewline
GSE95233 & GEO & 51 & 34 & 17 \tabularnewline
GSE54514 & GEO & 35 & 26 & 9 \tabularnewline
\midrule
Total & -- & 936 & 688 & 248 \tabularnewline
\bottomrule
\end{tabular}
\end{table}

\subsection{Outcome Definition}

Binary mortality classification was the selected prediction task. Splitting into two classes, survival was encoded as class 0 and death as class 1. Accounting for outcome annotation differentiations across studies, cohort-specific metadata was synchronized into a common binary endpoint before model development. Samples that were not accompanied by a specific mortality label were excluded during preprocessing.

\subsection{Transcriptomic Harmonization}

All cohorts were independently processed prior to any cross-cohort integration. Expression measurements were mapped to gene symbols using the corresponding platform annotations, duplicate probes mapping to the same gene were collapsed by mean expression and genes absent from one or more cohorts were removed to ensure that every retained feature was observed across the complete benchmark. The product of cohort intersection was a shared gene expression space containing 7,660 genes. To reduce platform-driven scale differences without borrowing information across cohorts, each sample was independently transformed by median/MAD robust scaling before cross-cohort modeling; no pooled batch-correction procedure was applied.

The produced harmonized gene space was anchored as the foundation for all subsequent molecular representations. Gene-level models operated directly on the shared expression matrix. Pathway-level models transformed the same matrix into 50 Hallmark pathway activity scores. Hybrid models combined both representations into a unified 7,710-feature set.

\section{Methodology}

\subsection{Molecular Representations}

Transcriptomic prognostic models can be constructed using molecular features at multiple levels of biological resolution. Gene-level representations preserve individual transcript measurements allowing them to maximize molecular detail. Pathway-level representations summarize coordinated biological programs using predefined gene sets. GSVA [5] and ssGSEA [6] have become widely adopted approaches for estimating pathway activity from gene-expression profiles because they reduce dimensionality while preserving biological interpretability. The Hallmark gene sets from MSigDB provide a curated collection of fifty coherent biological programs designed to minimize redundancy and represent canonical cellular processes [7].

Building upon these established approaches, three complementary transcriptomic representations were constructed using the harmonized common gene space described in Section III. The first representation consisted of the shared expression matrix containing 7,660 genes and preserved the complete transcriptomic resolution of the benchmark. The second representation transformed the same expression matrix into fifty Hallmark pathway activity scores using GSVA, compressing thousands of genes into biologically meaningful functional programs. Finally, a hybrid representation was generated by concatenating the gene-expression features with the corresponding pathway activity scores, allowing predictive models to simultaneously utilize fine-grained molecular information together with higher-level biological structure.

All three representations were derived from the identical harmonized transcriptomic data, ensuring that subsequent comparisons reflected differences in representation rather than differences in preprocessing or cohort composition. An overview of the full analytical pipeline is provided in Fig. 1.

\paperfig{1}{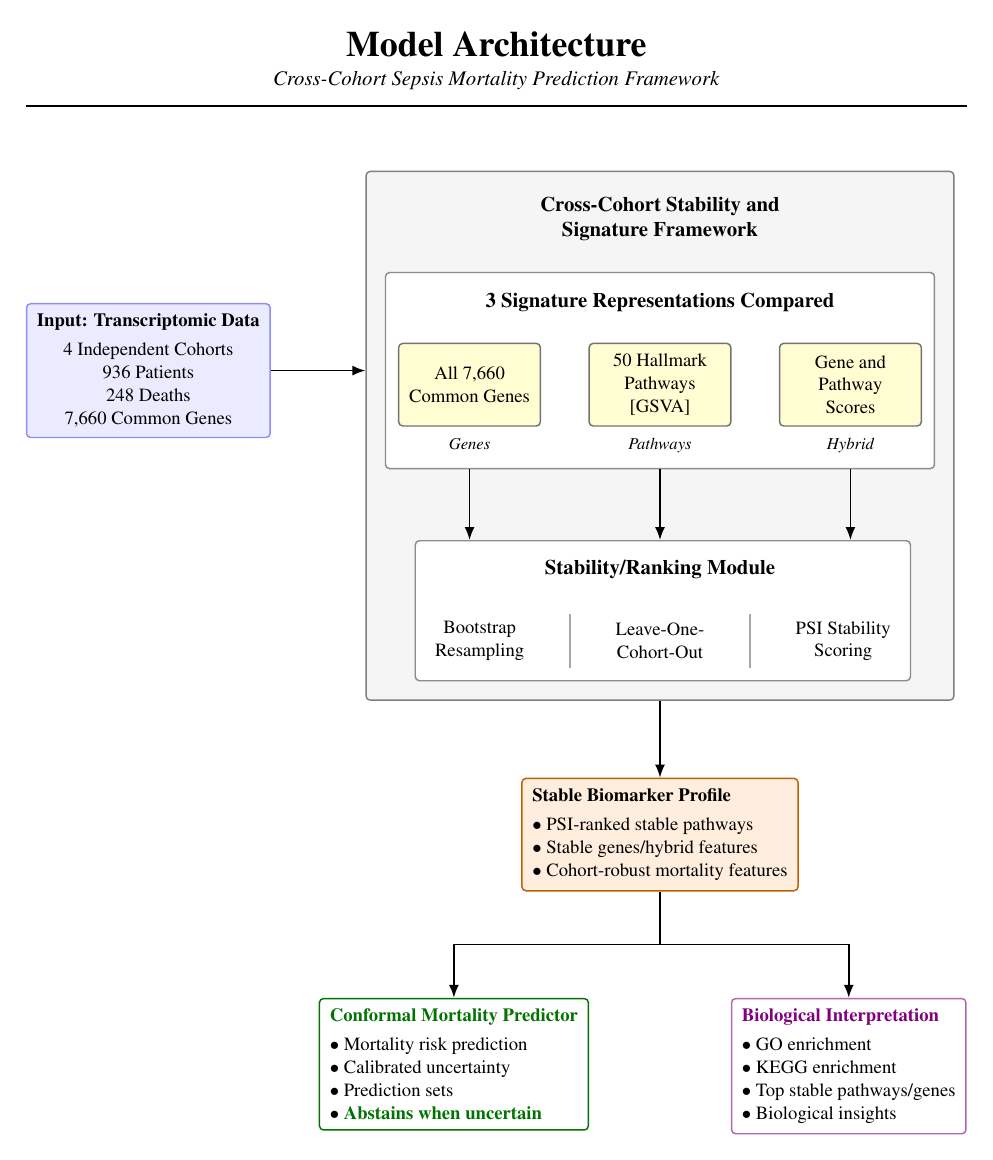}{Overview of the cross-cohort sepsis mortality prediction framework. Four independent transcriptomic cohorts are aligned into a common gene space, transformed into gene-level, pathway-level and hybrid representations, evaluated under leave-one-cohort-out validation and analyzed through stability ranking, conformal uncertainty estimation and biological enrichment.}{fig:pipeline}

\subsection{Pathway Stability Feature Index}

Feature selection within high-dimensional transcriptomic data is oftentimes sensitive to sampling variation, motivating the use of stability selection as a mechanism for identifying reproducible biomarkers [19]. Motivated by this principle, we introduce the Pathway Stability Index (PSI): a quantitative measure designed to identify transcriptomic signals that remain stable within the training cohorts rather than appearing only within a single bootstrap sample.

For each LOCO training fold, bootstrap resampling was performed exclusively on the three training cohorts. In bootstrap replicate $b$, an effect score was computed for each feature $f$ as the difference between mean expression among nonsurvivors and survivors, $e_{f,b}=\bar{x}_{f,b}^{(1)}-\bar{x}_{f,b}^{(0)}$. The top $K$ features by $|e_{f,b}|$ were selected, with $K=100$ for gene and hybrid representations and $K=20$ for pathway representations. Across $B=100$ bootstrap replicates, selection stability was defined as
\begin{equation}
S_f = \frac{1}{B}\sum_{b=1}^{B}\mathbf{1}(f\in A_b)
\end{equation}
where $A_b$ denotes the selected feature set in replicate $b$. Directional consistency was defined as
\begin{equation}
D_f = \left|\frac{\sum_{b=1}^{B}\mathbf{1}(f\in A_b)\operatorname{sign}(e_{f,b})}{\max(1,\sum_{b=1}^{B}\mathbf{1}(f\in A_b))}\right|
\end{equation}
The PSI score used for feature ranking was then computed as
\begin{equation}
\mathrm{PSI}(f) = \frac{1}{2}(S_f + D_f)
\label{eq:psi}
\end{equation}
Although motivated by pathway-level modeling, the same stability calculation was applied to gene, pathway and hybrid features. External held-out cohort labels were not used to compute PSI. Instead, external reproducibility was reported as a validation diagnostic after ranking by PSI, using agreement between the sign of the training-cohort effect and the sign of the held-out cohort effect. This separation was used to avoid circularity between stability ranking and external evaluation.

\subsection{Predictive Modeling Under Leave-One-Cohort-Out Evaluation}

Predictive performance was evaluated with LOCO validation in order to approximate real-world deployment to previously unseen hospitals. During each evaluation fold, models were trained using three cohorts while the remaining cohort served exclusively as an external test set. In contrast to random train-test splits or pooled cross-validation, LOCO evaluation prevents patients from the target institution from influencing model development and allows a substantially more realistic assessment of cross-cohort generalization.

Four supervised learning algorithms representing linear and ensemble modeling paradigms were evaluated: logistic regression, random forests [16], XGBoost [17] and LightGBM [18]. Logistic regression used standard scaling, balanced class weights and a maximum of 2,000 iterations. Random forests, XGBoost and LightGBM each used 400 estimators; tree ensembles were run with fixed random seeds and class-balancing where supported. Independent training using the same gene-level, pathway-level and hybrid transcriptomic representations was enforced across each classifier, and performance was quantified using AUROC, AUPRC and balanced accuracy.

All scaling, feature construction, feature ranking, model fitting, regularization selection, conformal calibration and stability estimation were performed using only the training cohorts within each LOCO fold. Held-out cohort labels were used only for final evaluation, except in the explicitly labeled oracle-calibration diagnostic. As an additional robustness check for logistic regression, the inverse regularization parameter $C$ was selected through inner leave-one-training-cohort-out validation over $C\in\{10,1,0.1,0.01,10^{-3},10^{-4},3\times10^{-5},10^{-5},3\times10^{-6},10^{-6},10^{-7}\}$ without accessing the true held-out cohort.

\subsection{Conformal Reliability and Selective Prediction}

While conventional discrimination metrics quantify ranking performance, they provide inadequate information regarding whether a model's confidence remains reliable after deployment. To quantify predictive uncertainty, we employed split conformal prediction: a model-agnostic framework that produces prediction sets with finite-sample coverage guarantees under exchangeability assumptions [8], [9]. Although these theoretical guarantees rely upon calibration and deployment data originating from the same underlying distribution, this assumption is frequently violated under cross-hospital transcriptomic deployment due to substantial cohort shift [10].

Conformal prediction was calibrated by splitting the three training cohorts in each LOCO fold into model-fitting and calibration subsets using a stratified 70/30 split. A balanced logistic-regression base estimator was fit on the model-fitting subset, conformalized on the calibration subset and evaluated on the held-out cohort at a nominal coverage target of 90\%. Realized coverage, coverage gap, mean prediction-set size and size-two prediction-set rate were recorded; size-two sets were interpreted as abstentions because the conformal predictor could not exclude either class. For diagnostic purposes, pooled LOCO calibration, cohort-specific calibration and oracle target-cohort calibration were additionally compared to distinguish failures arising from distribution shift from limitations of the conformal framework itself.

Selective prediction was also explored through confidence-threshold abstention. Models were given permission to abstain on uncertain samples, allowing the relationship between retained-case performance and prediction coverage to be quantified. Calibration quality was assessed using the Brier score alongside the Expected Calibration Error (ECE), providing complementary measures of probabilistic reliability.

\subsection{Biological Interpretation}

By itself, statistical reproducibility does not guarantee biological relevance, and realizing this, transcriptomic features with high PSI values were further investigated using functional enrichment analysis. Gene symbols corresponding to stable features were converted to Entrez identifiers before Gene Ontology (GO) and Kyoto Encyclopedia of Genes and Genomes (KEGG) enrichment analyses were performed. The goal of these analyses was to determine whether transcriptomic signals identified across cohorts corresponded to established biological processes associated with sepsis including immune activation, inflammatory signaling, host-defense mechanisms, leukocyte function and antimicrobial responses.

By combining pathway-level representation learning, stability-based feature selection, rigorous cross-cohort evaluation, conformal uncertainty estimation and biological interpretation within a single pipeline, the proposed methodology enables representation choice to be evaluated not just with respect to predictive discrimination, but also with respect to robustness, calibration, transferability and biological coherence.

\section{Experimental Results}

\subsection{Cross-Cohort Predictive Performance}

The first evaluation of predictive performance was done under the four-cohort LOCO setting. This evaluation condition is purposely more difficult compared to pooled or random-split validation because each model is tested on a completely unseen cohort and because of this the observed performance reflects cross-cohort transfer rather than interpolation within a mixed dataset.

\begin{table}[t]
\centering
\caption{Leave-one-cohort-out model performance summary across molecular representations.}
\label{tab:sweep}
\scriptsize
\setlength{\tabcolsep}{3.2pt}
\renewcommand{\arraystretch}{1.08}
\begin{tabular}{@{}llccc@{}}
\toprule
\textbf{Rep.} & \textbf{Model} & \textbf{AUROC} & \textbf{AUPRC} & \textbf{Bal. Acc.} \tabularnewline
\midrule
Genes & LogReg & 0.547 & 0.403 & 0.466 \tabularnewline
Genes & RF & 0.529 & 0.327 & 0.500 \tabularnewline
Genes & XGBoost & 0.548 & 0.321 & 0.504 \tabularnewline
Genes & LightGBM & 0.604 & \textbf{0.428} & 0.522 \tabularnewline
\addlinespace
Pathways & LogReg & 0.564 & 0.343 & 0.515 \tabularnewline
Pathways & RF & 0.506 & 0.311 & 0.495 \tabularnewline
Pathways & XGBoost & 0.540 & 0.317 & \textbf{0.536} \tabularnewline
Pathways & LightGBM & 0.512 & 0.286 & 0.474 \tabularnewline
\addlinespace
Hybrid & LogReg & 0.548 & 0.406 & 0.466 \tabularnewline
Hybrid & RF & 0.601 & 0.376 & 0.500 \tabularnewline
Hybrid & XGBoost & 0.582 & 0.394 & 0.529 \tabularnewline
Hybrid & LightGBM & \textbf{0.609} & 0.426 & 0.519 \tabularnewline
\bottomrule
\end{tabular}
\end{table}

\begin{table}[t]
\centering
\caption{Primary logistic-regression reliability summary across LOCO folds. Coverage uses a nominal target of 0.90; $|\Delta_{cov}|$ is the mean absolute coverage gap.}
\label{tab:reliability}
\scriptsize
\setlength{\tabcolsep}{2.5pt}
\renewcommand{\arraystretch}{1.05}
\begin{tabular}{@{}lcccccc@{}}
\toprule
\textbf{Rep.} & \textbf{AUROC} & \textbf{AUPRC} & \textbf{Cov.} & \textbf{$|\Delta_{cov}|$} & \textbf{ECE} & \textbf{Brier} \tabularnewline
\midrule
Genes & 0.548 & 0.405 & 0.564 & 0.366 & 0.589 & 0.578 \tabularnewline
Pathways & 0.563 & 0.343 & \textbf{0.910} & \textbf{0.026} & \textbf{0.211} & \textbf{0.256} \tabularnewline
Hybrid & 0.549 & 0.405 & 0.564 & 0.367 & 0.595 & 0.585 \tabularnewline
\bottomrule
\end{tabular}
\end{table}

Within the entire model sweep it was observed that predictive discrimination was dependent on both the molecular representation and the learning algorithm. Gene-level and hybrid representations generally achieved the strongest raw discrimination under realistic cross-cohort evaluation. This is seen as the strongest overall AUROC was achieved by the hybrid representation combined with LightGBM (0.609), while the highest mean AUPRC was obtained using the gene-level representation with LightGBM (0.428), closely followed by the hybrid representation with the same classifier (0.426). In the leakage-free nested-regularization sensitivity analysis, gene and hybrid logistic-regression AUROC improved from approximately 0.548--0.549 under the default regularization setting to approximately 0.608--0.607 after inner-cohort selection of $C$, whereas pathway logistic-regression AUROC decreased from 0.563 to 0.514; this indicates that high-dimensional representations are more hyperparameter-sensitive but can be tuned without accessing the target cohort.

Pathway-level models remained competitive across multiple learning algorithms despite only using 50 Hallmark pathway scores and not the 7,660 shared gene feature space. For instance, pathway-level logistic regression achieved a mean AUPRC of 0.343, exceeding the performance of both gene-level Random Forest (0.327) and gene-level XGBoost (0.321). Furthermore, pathway-level XGBoost produced the highest balanced accuracy among all evaluated representation-model combinations (0.536), suggesting that pathway aggregation preserved clinically relevant discriminative information despite substantial dimensionality reduction.

\paperfig{0.86}{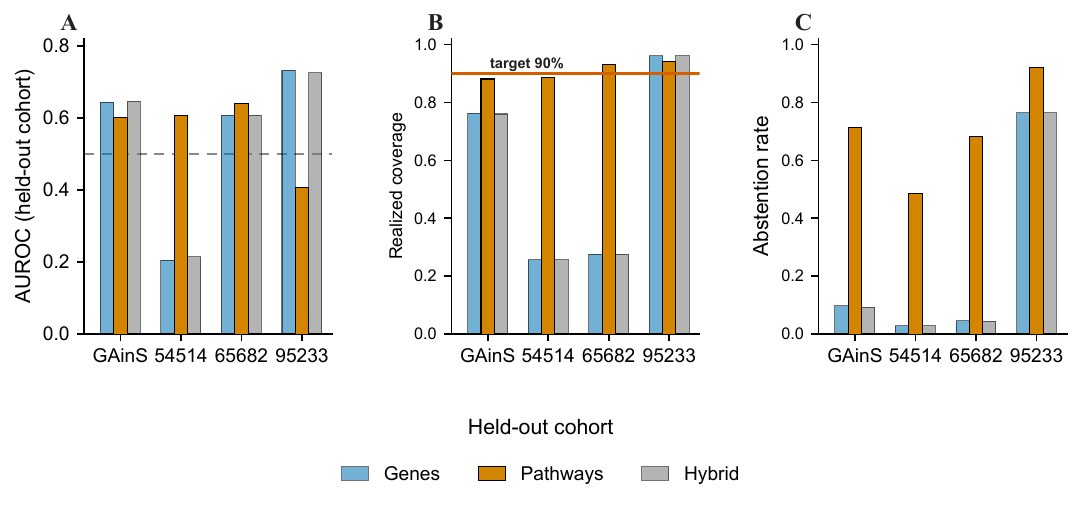}{Cross-cohort model behavior across transcriptomic representations. Panel A shows AUROC on held-out cohorts under leave-one-cohort-out validation. The dashed line marks chance-level AUROC. Panel B displays realized conformal coverage under a nominal target of 0.90. Panel C exhibits abstention rate across held-out cohorts. Genes, pathways, and hybrid representations are shown in blue, orange and gray, respectively.}{fig:model_performance}

The model performance chart displayed in Fig.~\ref{fig:model_performance} demonstrates that no singular representation metric reigned supreme across every evaluation metric. Gene-level and hybrid representations generally achieved the highest discrimination, whereas pathway-level models exhibited a decent reduction in AUROC and AUPRC while remaining competitive across the four excluded cohorts. The trade-off indicates that pathway aggregation sacrifices little predictive performance despite reducing the feature space by more than two orders of magnitude. Seeing that the differences in discrimination between representations were comparatively modest, the remaining analyses focus on characteristics uncaptured by AUROC and AUPRC, including robustness to model choice, uncertainty reliability, calibration, selective prediction and biological reproducibility.

\subsection{Reliability Under Cross-Cohort Shift}

Predictive discrimination is a tool for estimating ranking performance but it cannot indicate whether a model's confidence remains reliable after deployment to an unseen hospital. This led us to evaluate split conformal prediction under the leave-one-cohort-out setting by comparing realized coverage against the nominal 90\% target.

\paperfig{0.88}{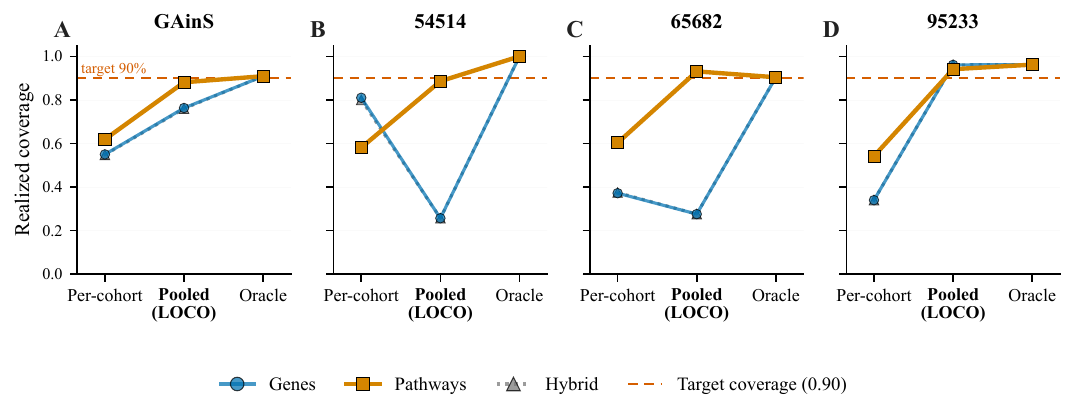}{Comparison of conformal calibration strategies across held-out cohorts. Each panel corresponds to one external test cohort and compares per-cohort calibration, pooled leave-one-cohort-out calibration and oracle target-cohort calibration. The dashed line indicates a coverage target of 0.90.}{fig:calibration_strategy}

Conformal coverage varied substantially and depended on molecular representation and calibration strategy when examined across the four held-out cohorts. High predictive performance alone did not guarantee trustworthy uncertainty estimates. As summarized in Table~\ref{tab:reliability}, pathway-level logistic regression maintained mean realized coverage close to the nominal target (0.910, mean absolute coverage gap 0.026), whereas gene-level and hybrid models were characterized by substantial undercoverage despite competitive discrimination metrics.

Oracle calibration functioned as a diagnostic experiment and when calibration was performed using target-cohort labels, realized coverage approached the nominal level across representations as conformal prediction itself remained capable of achieving valid coverage when calibration and deployment distributions were aligned. The primary source of reliability degradation was therefore the distribution shift between hospitals rather than an inherent limitation of the conformal framework. An important distinction between predictive discrimination and predictive reliability is thus identified in that a model may correctly rank patients according to risk while simultaneously providing confidence estimates that become unreliable when transferred to a new institution.

\subsection{Calibration and Selective Prediction}

Well-calibrated confidence estimates is one of the foremost requirements for reliable deployment alongside accurate ranking. To incorporate this, we quantified probabilistic calibration using the Brier score and Expected Calibration Error (ECE) and subsequently evaluated confidence-threshold abstention to determine whether selectively withholding uncertain predictions improved retained-case performance.

\paperfig{0.84}{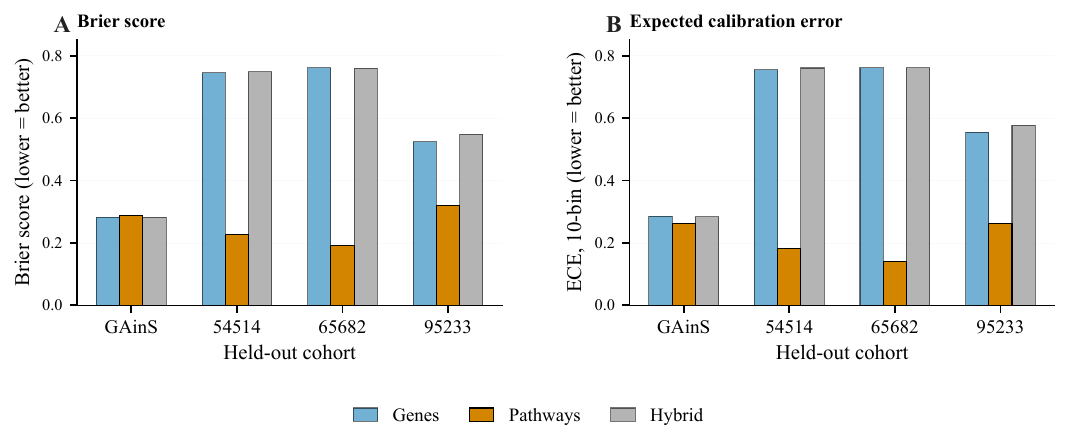}{Probability calibration across transcriptomic representations. (A) Brier score across held-out cohorts. (B) Expected calibration error computed using 10 bins. Lower values indicate better probabilistic calibration in both panels.}{fig:calibration_metrics}

Gene-level and hybrid representations generally exhibited stronger discrimination but also larger calibration error under several held-out cohorts. On the other hand, pathway representations produced slightly lower discrimination while maintaining lower average ECE and Brier score in the logistic-regression reliability analysis, indicating that pathway aggregation generated probability estimates that were more consistent with observed outcomes.

\paperfig{0.84}{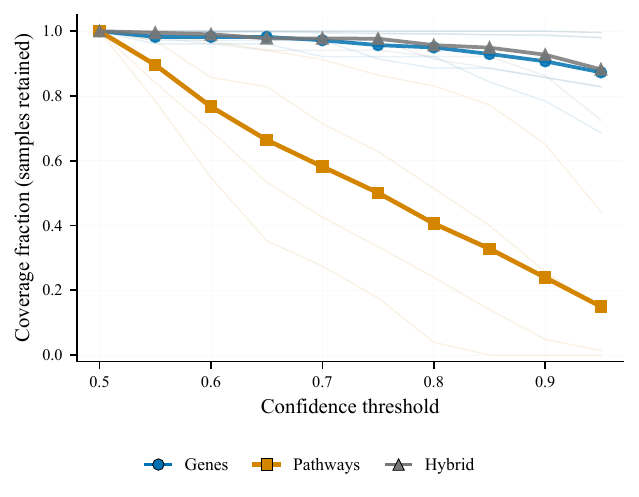}{Confidence-threshold abstention behavior across transcriptomic representations. The x-axis shows the confidence threshold required to issue a prediction, and the y-axis shows the fraction of samples retained after abstention. Faint lines show cohort-level traces, while bold lines summarize representation-level behavior.}{fig:confidence_abstention}

Retained-case performance did not improve uniformly as uncertain predictions were removed, demonstrated by confidence-threshold abstention. Selective prediction occasionally was able to increase AUPRC on retained patients yet the benefit depended on whether underlying confidence estimates held reliable when exposed to cohort shift. For this reason, we formulated that abstention should be interpreted as a supplemental deployment strategy rather than a replacement for accurate uncertainty estimation.

\subsection{Pathway Stability Index Identifies Transferable Transcriptomic Signals}

Next in our evaluation was exploring whether the proposed Pathway Stability Index was able to pinpoint features that remained reproducible across independent cohorts. Differing from conventional feature importance measures, PSI was designed to prioritize signals that were consistently selected during bootstrap resampling. The metric also preserved their direction of association with mortality and reproduced similar behavior under external validation.

\paperfig{0.84}{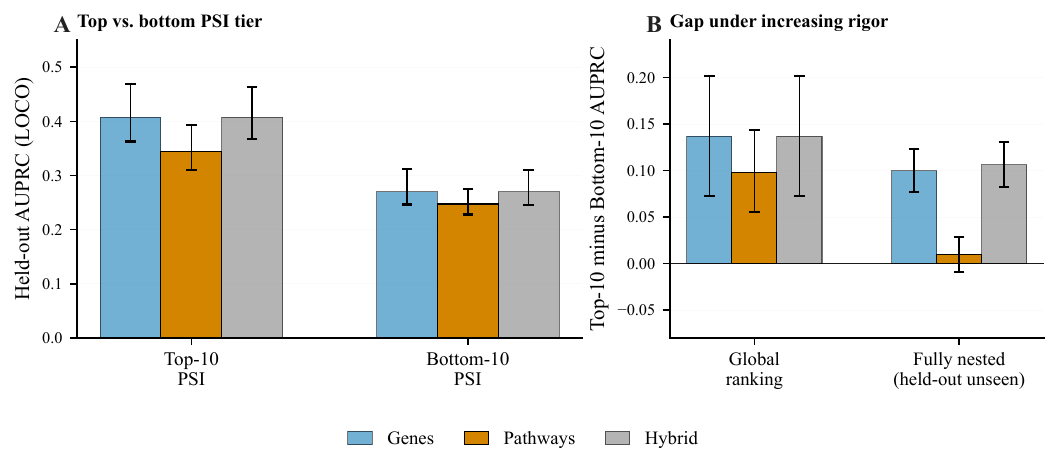}{Validation of the Pathway Stability Index under held-out cohort evaluation. (A) Held-out AUPRC for top-10 and bottom-10 PSI feature tiers. (B) Difference in AUPRC between top-10 and bottom-10 PSI tiers under global ranking and fully nested held-out evaluation. Error bars indicate uncertainty across validation splits or bootstrap resampling.}{fig:psi_validation}

High-PSI pathways, as defined by~\eqref{eq:psi}, had considerably greater external reproducibility compared to low-PSI pathways and also simultaneously exhibited higher selection frequency and directional consistency. PSI successfully separated biological signals that generalized across independent cohorts from associations that were unique to individual datasets. High PSI scores reflected consistency across studies rather than predictive performance within any one cohort. Some of the strongest single-cohort predictors ranked lower because they could not be reproduced in other datasets.

\subsection{Stable Transcriptomic Signatures Recover Immune and Host-Defense Biology}

PSI-ranked genes were also subjected to Gene Ontology and KEGG enrichment analysis to determine whether highly stable transcriptomic features corresponded to established sepsis pathophysiology mechanisms. Representative enriched Gene Ontology biological-process terms and KEGG pathways are summarized in Table~\ref{tab:enrichment}.

\begin{table*}[!t]
\centering
{\scriptsize
\caption{Functional enrichment of high-stability (high-PSI) transcriptomic features. Representative top Gene Ontology biological-process terms and KEGG pathways are ranked by Benjamini--Hochberg adjusted $p$-value (FDR). \textit{Genes}: number of stable features annotated to the term; \textit{Fold Enr.}: fold enrichment over the genomic background.}
\label{tab:enrichment}
\setlength{\tabcolsep}{4pt}
\renewcommand{\arraystretch}{0.94}
\begin{tabular}{@{}llccc@{}}
\toprule
\textbf{ID} & \textbf{Term} & \textbf{Genes} & \textbf{Fold Enr.} & \textbf{FDR ($q$)} \tabularnewline
\midrule
\multicolumn{5}{@{}l}{\textbf{Gene Ontology (Biological Process)}} \tabularnewline
GO:0006959 & humoral immune response & 23 & 7.8 & $2.3\times10^{-11}$ \tabularnewline
GO:0009617 & response to bacterium & 39 & 4.1 & $3.7\times10^{-11}$ \tabularnewline
GO:0042742 & defense response to bacterium & 24 & 5.9 & $1.7\times10^{-9}$ \tabularnewline
GO:0019730 & antimicrobial humoral response & 14 & 9.7 & $4.3\times10^{-8}$ \tabularnewline
GO:0007159 & leukocyte cell-cell adhesion & 30 & 3.5 & $8.3\times10^{-7}$ \tabularnewline
GO:0006954 & inflammatory response & 40 & 2.7 & $1.9\times10^{-6}$ \tabularnewline
\addlinespace
\multicolumn{5}{@{}l}{\textbf{KEGG Pathways}} \tabularnewline
hsa05150 & Staphylococcus aureus infection & 16 & 9.7 & $3.2\times10^{-10}$ \tabularnewline
hsa04145 & Phagocytosis & 30 & 4.2 & $1.1\times10^{-9}$ \tabularnewline
hsa04640 & Hematopoietic cell lineage & 13 & 5.3 & $3.9\times10^{-5}$ \tabularnewline
hsa05140 & Leishmaniasis & 11 & 5.1 & $2.2\times10^{-4}$ \tabularnewline
\bottomrule
\end{tabular}
}
\vspace{-2mm}
\end{table*}

Immune and host-defense processes were the highest ranking enrichment terms. Representative examples include antibacterial responses, antimicrobial humoral immunity, leukocyte activation, inflammatory signaling and antigen presentation. These medical themes support the biological plausibility of the stable transcriptomic signals identified by PSI and are consistent and aligned with dysregulated host immune response that characterizes sepsis.

Although several highly ranked features mapped directly to immune programs, not every statistically stable signal had an immediately obvious mechanistic relationship to sepsis. These findings indicate PSI to be most useful when paired with functional enrichment and domain knowledge.

Overall, the proposed stability framework preferentially identifies coherent host-response programs rather than isolated cohort-specific biomarkers. Pathway-level stability, in consequence, provides a useful bridge between predictive machine learning models and biologically interpretable mechanisms that may ultimately support more reliable deployment of transcriptomic prognostic models in clinical settings.

\section{Discussion}

The key result of this study is that the type of molecular representation used in a model can impact performance beyond discrimination alone. While the gene-level and hybrid transcriptomic representations most often resulted in the highest AUROC and AUPRC values when using leave-one-cohort-out evaluation, those values were not as consistent for generating reliable uncertainty estimates when transferred to new cohorts. In general, pathway-level representations generated somewhat poorer discrimination on average; however they tended to be more stable across different machine learning algorithms, provided better conformal behavior when changing from one cohort to another and had transcriptomic signatures that could have been supported by the biology. This provides evidence that using discrimination measures alone to evaluate biomedical models intended for deployment can give a limited view of how reliably those models will perform in practice.

A possible explanation for the observed outcomes lies in the role of pathway aggregation acting as a form of biological regularization. By reducing thousands of transcriptomic measurements down to coordinated functional programs through pathway scores, pathway aggregation provides a method of reducing the dimensionality of data, while maintaining the core biological responses of the host. While this may result in losing molecular detail, it appears that the pathway representations are less susceptible to model selection bias and more resistant to variations in cohorts, thus allowing for greater consistency in generalizing the models across multiple independent clinical studies.

Predictive performance analysis in conjunction with uncertainty analysis showed that discrimination is insufficient to determine whether a model is clinically deployable. Models that were highly discriminating exhibited an increasing level of confidence as their performance decreased on external data. Conversely, models at the pathway level often represented uncertainty by either producing larger conformal prediction sets or selectively withholding predictions. Abstaining from making a prediction is not a substitute for properly calibrated predictive output; however, withholding predictions when there is a large distribution shift in the test cohort is preferable to consistently producing confident but incorrect probability estimates.

There are a number of limitations to this research. While the benchmark contains four separate cohorts consisting of 936 patients, there remains considerable heterogeneity among cohort sizes and transcriptomic platforms, including one small held-out cohort with 35 patients and 9 deaths; therefore, fold-level estimates should not be interpreted as definitive clinical performance guarantees. All analyses were also conducted retrospectively using publicly available data. Conformal guarantees decrease as the extent of distribution shift increases. Additionally, while PSI can identify biologically plausible and externally validated signals based upon results in this benchmark, additional validation will be required to ensure that PSI performs similarly across larger and more diverse datasets. Future research should validate this framework in prospective multicenter cohorts, explore other pathway resources and graph-based biological representations, and incorporate distribution-shift detection before prediction.

\section{Conclusion}

This study presented and evaluated a multifaceted framework for evaluating transcriptomic sepsis mortality prediction under realistic cross-cohort deployment. By utilizing four independent cohorts composed of 936 patients total, we systematically compared gene-level, pathway-level, and hybrid transcriptomic representations under leave-one-cohort-out validation. This was done alongside extending evaluation beyond conventional discrimination metrics to include conformal reliability, calibration, selective prediction, feature stability, and biological interpretation.

Throughout the benchmark, gene-level and hybrid representations frequently achieved the strongest predictive discrimination. Yet, pathway-level representations demonstrated greater robustness across several learning algorithms. In addition, these representations featured more reliable uncertainty behavior under cohort shift and stable transcriptomic signatures that aligned with established immune and host-defense processes. The findings denote that maximizing AUROC or AUPRC alone is insufficient when developing models intended for deployment across independent clinical settings.

Broadly, this study reveals that molecular representation influences not only predictive performance but also the reliability, interpretability, and transferability of transcriptomic prognostic models.

\end{document}